\documentclass[12pt,aps,prd,preprint,nofootinbib]{revtex4}   

\usepackage{amsmath}    
\usepackage{amsfonts}   
\usepackage{amssymb}
\usepackage{graphicx}   

\begin{document}

\title{Schwarzschild and de Sitter solutions from the argument by Lenz and Sommerfeld}
\author{R. R. Cuzinatto}
 \email{cuzinatto@unifal-mg.edu.br}  
 \affiliation{Instituto de Ci\^{e}ncia e Tecnologia, Universidade Federal de Alfenas, Campus Po\c{c}os de
 Caldas, Rodovia Po\c{c}os/S\~{a}o Paulo (BR 267), km 533, 11999, CEP 37701-970, Po\c{c}os de Caldas, MG, Brazil}
\author{B. M. Pimentel}
 \email{pimentel@ift.unesp.br}
 \affiliation{Instituto de F\'{\i}sica Te\'{o}rica, UNESP - S\~{a}o Paulo State University, P.O. Box 70532-2, CEP 01156-970, S\~{a}o Paulo, SP, Brazil}
\author{P. J. Pompeia}
 \email{pedropjp@ifi.cta.br}
 \affiliation{Departamento de Ci\^{e}ncia e Tecnologia Aeroespacial, Instituto de Fomento e Coordena\c{c}\~{a}o Industrial, Pra\c{c}a Mal. Eduardo Gomes,
 50,\\ CEP\ 12228-901, S\~{a}o Jos\'{e} dos Campos, SP, Brazil}

\begin{abstract}
The Lenz-Sommerfeld argument allows an ingenious and simple
derivation of the Schwarzschild solution of Einstein equations of
general relativity. In this paper, we use the same reasoning to
construct the de Sitter line element.
\end{abstract}

\maketitle

\section{General comments on gravity \label{Sec-Intro}}

Einstein's first step towards the general theory of relativity was a
\textit{gedanken} experiment with an elevator. He considered an
observer locked in a box at rest in the Earth's gravitational field
performing simple experiments, such as throwing objects and timing
the period of a pendulum. This person would see the projectile
describing a parabolic trajectory
downward \cite{Halliday}%
\begin{equation}
y=\left(  \tan\theta_{0}\right)  x-\left(  \frac{g}{2v_{0}^{2}\cos^{2}%
\theta_{0}}\right)  x^{2}~,\label{parabola}%
\end{equation}
where $g$\ is the magnitude of the Earth's gravitational
acceleration, $y$ is the vertical distance, $x$ is the horizontal
range and $v_{0}$\ is the velocity upon launch at an angle
$\theta_{0}$ with the horizontal direction. Also, he would verify
that the period $T$ of the simple pendulum measured with
the clock could also be obtained by calculating%
\begin{equation}
T=2\pi\sqrt{\frac{L}{g}}~;\label{T}%
\end{equation}
$L$ is the length of the pendulum. Then Einstein imagined this same
box, observer and experiments put in the interstellar space freed
from any gravitational influence, accelerated upwards with magnitude
$g$. As he reported later,\footnote{As stated in Ref.
\cite{Halliday}, Chapter 15, page 430.} Einstein was astonished with
the fact that the results of the experiments carried by the observer
in this new situation would be exactly the same: the thrown object
would still follow a parabolic path -- according to Eq.
(\ref{parabola})\ -- and the period of the pendulum would not change
-- Eq. (\ref{T}) would be valid yet.

Instead, if we consider the elevator free falling in the Earth's
gravitational field, the observer (of mass $m$, say) would cease to
press the floor with a force of magnitude $mg$: in the non-inertial
reference system attached to the elevator, he is\ \textquotedblleft
weightless\textquotedblright, in exactly the same way as the space
shuttle astronauts float along with their equipment while they orbit
Earth. In the free falling reference system, $g=0$ and the parabolic
trajectory of ballistic motion degenerates into a straight line,
once
Eq. (\ref{parabola}) reduces to%
\begin{equation}
y=\left(  \tan\theta_{0}\right)  ~x~,\label{straight line}%
\end{equation}
Moreover, the period of oscillation of the pendulum becomes
infinity: $T\rightarrow\infty$ as $g\rightarrow0$ in accordance with
Eq. (\ref{T}).

The reason for Einstein's amusement is born in a hypothesis hidden
in Eq. (\ref{T}), namely the \emph{equivalence between inertial and
gravitational
mass}:%
\begin{equation}
m_{i}=m_{g}~.\label{mi=mg}%
\end{equation}
Should the coefficient that measures the response to any kind of
force $m_{i} $\ (the mass in Newton's second law of motion,
$\mathbf{F}=m_{i}\mathbf{a}$) be any different from the inertia
coefficient to gravitation $m_{g}$ (the mass in the definition of
weight,\textbf{\ }$\mathbf{W}=m_{g}\mathbf{g}$), the
period of a pendulum would be given by%
\begin{equation}
T=2\pi\sqrt{\frac{m_{i}L}{m_{g}g}}\label{T(mi/mg)}%
\end{equation}
rather than by Eq. (\ref{T}). Experiments of extremely high
accuracy, such as those by R. E\"{o}tv\"{o}s with the torsion
balance, guarantee that Eq. (\ref{mi=mg}) holds up to a precision of
1 part in $10^{12}$.

The second move in Einstein's reasoning was to associate gravitation
and inertia to world curvature. This way, physics was brought from
the flat spacetime structure of Minkowski's line element of special
relativity to the curved geometry studied by Gauss, Riemmann and
Levi-Civita, among others.\ According to this innovative idea,
gravity would not be described by a vector field related to Newton's
law,
\begin{equation}
\mathbf{F}_{N}\left(  r\right)  =-\frac{GmM}{r^{2}}~\hat{r}%
\;,\label{Newton force}%
\end{equation}
for the attractive gravitational force $\mathbf{F}_{N}$\ acting on
bodies of masses $m$ and $M$ separated by a distance $r$ along the
direction connecting the centers of mass. In Einstein's theory of
gravitation, the field has a tensorial character: it corresponds to
a $4\times4$\ matrix-like object with ten independent components,
$g_{\mu\nu}$, defining the arc length $s$\ and
infinitesimal distances%
\begin{equation}
ds^{2}=g_{\mu\nu}dx^{\mu}dx^{\nu}\label{ds2}%
\end{equation}
on the curved manifold of experience. Following Mach's ideas,
Einstein proposed that the presence of mass sets the stage for the
events to take place: $g_{\mu\nu}$\ is a solution of \cite{Ad75}
\begin{equation}
R_{\mu\nu}-\frac{1}{2}g_{\mu\nu}R+\Lambda g_{\mu\nu}=-\kappa
T_{\mu\nu
}~.\label{Einstein Eq.}%
\end{equation}
The distribution of mass is modeled by the stress-energy tensor
$T_{\mu\nu}$; the Ricci tensor $R_{\mu\nu}$ and scalar curvature
$R$\ are functions of (the derivatives of) $g_{\mu\nu}$. The factor
$\kappa$\ is basically the Newton's constant $G$ appearing in Eq.
(\ref{Newton force})\ divided by (the fourth power of) the speed of
light $c$: $\kappa=8\pi G/c^{4}$. The cosmological constant
$\Lambda$ was introduced by Einstein so he could derive, out of his
theory, a static spherical model for the universe \cite{To87}, which
matched the beliefs of that epoch.

Nowadays, $\Lambda$ is essential to physical cosmology for the
opposite reason, once it allows for the ever expanding cosmos. The
so called de Sitter model, named after the man who built it in 1917
\cite{dS17}, is a solution of (\ref{Einstein Eq.}) that predicts an
increasingly accelerated recession of the galaxies. The de Sitter
interval takes the form
\begin{equation}
ds^{2}=\frac{dr^{2}}{1-\frac{\Lambda r^{2}}{3}}+r^{2}\left(  d\theta^{2}%
+\sin^{2}\theta d\phi^{2}\right)  -\left(  1-\frac{\Lambda
r^{2}}{3}\right)
c^{2}dt^{2}\label{deSitter}%
\end{equation}
in the static spherical coordinates $\left(  r,\theta,\phi\right)
$. In comoving coordinates, the de Sitter line element is written as
\begin{equation}
ds^{2}=a_{0}e^{\sqrt{\frac{4\Lambda c^{2}}{3}}\left(  t-t_{0}\right)
}\left[ dr^{2}+r^{2}\left(  d\theta^{2}+\sin^{2}\theta
d\phi^{2}\right)  \right]
-c^{2}dt^{2}\label{dS-Friedmann}%
\end{equation}
where the radial coordinate $r$ is dimensionless and the physical
distance is given by the product of $r$ by the scale factor $a\left(
t\right)  $ given by
(see e.g. \cite{Al06}):%
\begin{equation}
a\left(  t\right)  =a_{0}e^{\sqrt{\frac{\Lambda c^{2}}{3}}\left(
t-t_{0}\right)  }\label{a dS}%
\end{equation}
($a_{0}=a\left(  t_{0}\right)  $ is the initial condition set for
the scale factor\footnote{Actually, $t_{0}$ usually represents the
present-day value of the cosmic time.}), which is equivalent to say
that it is $a\left(  t\right) $\ that carries the dimension of
length. Observing the form of the dependence of $a$ on $t$ in Eq.
(\ref{a dS}),\ it is clear that the de Sitter solution describes the
accelerated expanding universe.

The scale factor (\ref{a dS})\ can also be determined using the
Friedmann
equations%
\begin{align}
& \left.  \left(  \frac{\dot{a}}{a}\right)  ^{2}=\frac{8\pi
G}{3}\rho
+\frac{\Lambda c^{2}}{3}-\frac{kc^{2}}{a^{2}}\right. \label{a dot}\\
& \left.  \frac{\ddot{a}}{a}=\frac{\Lambda c^{2}}{3}-\frac{4\pi
G}{3}\left(
\rho+\frac{3p}{c^{2}}\right)  \right. \label{a ddot}%
\end{align}
for $a\left(  t\right)  $ appearing in a general metric of the form%
\begin{equation}
ds^{2}=-c^{2}dt^{2}+a^{2}\left(  t\right)  \left[  \frac{dr^{2}}{1-kr^{2}%
}+r^{2}\left(  d\theta^{2}+\sin^{2}\theta d\phi^{2}\right)  \right]
\label{Friedmann}%
\end{equation}
where $k$ is a number describing a space section with \emph{(i)}
flat geometry for $k=0$; \emph{(ii)} spherical geometry if $k=+1$;
and \emph{(iii)} hyperbolic geometry for $k=-1$. In the pair (\ref{a
dot}-\ref{a ddot}), $\rho $\ and $p$\ are the mass density and the
pressure of the massive content in our model of universe. The dot
means derivative with respect to the cosmic time:
$\dot{a}=\frac{da}{dt}$.

The second Friedmann equation\ enables us to give an alternative
interpretation to the effect of the cosmological constant $\Lambda$
\cite{Al06}. For this purpose, consider a region limited by a sphere
of physical radius $a\left(  t\right)  r_{0}$ (with a constant
$r_{0}$). Then,
from Eq. (\ref{a ddot}),%
\begin{equation}
\frac{d^{2}\left(  ar_{0}\right)  }{dt^{2}}=\frac{\Lambda
c^{2}}{3}\left(
ar_{0}\right)  -\frac{GM}{\left(  ar_{0}\right)  ^{2}}\label{Sphere Eq.}%
\end{equation}
where we have identified%
\begin{equation}
M=\frac{4\pi}{3}\left(  \rho+\frac{3p}{c^{2}}\right)  \left(
ar_{0}\right)
^{3}~,\label{M}%
\end{equation}
the \textit{total mass} within the sphere: amongst all the energetic
density $\left(  \rho+\frac{3p}{c^{2}}\right)  $\ homogeneously and
isotropically distributed in the universe, we select the sector
inside the volume $\frac{4\pi}{3}\left(  ar_{0}\right)  ^{3}$ of the
sphere under consideration. Multiplying Eq. (\ref{Sphere Eq.})\ by
the mass $m$ of a test particle, we get
a force equation%
\begin{equation}
F=F_{\Lambda}+F_{N}\label{Force Eq.}%
\end{equation}
whose second term is the Newtonian law of gravity (\ref{Newton
force}),
whereas the first term on the r.h.s,%
\begin{equation}
\mathbf{F}_{\Lambda}\left(  r\right)  =m\frac{\Lambda
c^{2}}{3}r~\hat
{r}\;,\label{dS force}%
\end{equation}
can be interpreted as a repulsive force due to
$\Lambda$.\footnote{The interpretation of the repulsive effect of
$\Lambda$\ in terms of a force is not de Sitter's contribution, and
actually it is not strictly meaningful in the context of general
relativity. In Einstein theory, the concept of force is abandoned in
favor of the curvature of the spacetime. However, as one comes to
$\mathbf{F}_{\Lambda}\left(  r\right)  $\ by taking the Friedmann
equations as the first step of the derivation, the interpretation is
at least in agreement with the results from the general theory of
gravity.} In this approach, the force $\mathbf{F}_{\Lambda}$\ would
be the one driving the expansion of the universe. And the recent
observations \cite{WMAP,SNIa}\ favor this accelerated expanding
scenario.

Einstein's theory of gravity is not only able to describe phenomena
explained by the Newtonian approach but also successfully addresses\
the observational puzzle of the perihelion shift of Mercury
\cite{Sa85}. It also predicts that light rays coming from distant
stars should be deviated from straight trajectories when passing by
the Sun \cite{Gasperini09}, a fact that was verified by Eddington
\cite{Eddington}.

The new theory calls for a philosophical change in the the way we
interpret gravity, and it comes along with a new set of
sophisticated mathematical tools. Schwarszchild had to integrate the
set of partial coupled differential equations (\ref{Einstein Eq.})
in order to obtain the line element for the spacetime surrounding a
massive body of the kin of a planet or a star (of mass $M$). His
solution, found as early as 1916, reads \cite{Schw16}
\begin{equation}
ds^{2}=\frac{dr^{2}}{1-\frac{2\alpha}{r}}+r^{2}\left(
d\theta^{2}+\sin
^{2}\theta d\phi^{2}\right)  -\left(  1-\frac{2\alpha}{r}\right)  c^{2}%
dt^{2}\label{Schwarzschild}%
\end{equation}
with $\alpha$ standing for the geometrical mass of the source%
\begin{equation}
\alpha=\frac{GM}{c^{2}}~.\label{geometrical mass}%
\end{equation}

To solve Einstein's equations given a certain distribution of matter
and energy can be an extremely laborious task, if possible at all.
Lenz developed an alternative method for deriving the Schwarzschild
solution. He did not publish his result, as far as these authors are
concerned, but he communicated his argument to Sommerfeld in 1944,
which reproduced its general lines in his classic treatise on
electrodynamics \cite{So52}. Lenz's approach combines Newton's
gravitation law with Einstein's special theory of relativity to get
a non-pseudo-Euclidean spacetime. In this paper we will revise this
reasoning (Section \ref{Sec-Lenz}), showing its solid grounds and
the efficiency of this approximative technique. We will also\ extend
it to produce the de Sitter solution (Section \ref{Sec-dS}).

\section{The argument by Lenz and the Schwarzschild interval \label{Sec-Lenz}}

As Lenz's reasoning is based on special theory of relativity, it is
appropriate to revisit the concepts of proper time and proper
length. Special relativity rises from two postulates: (1) the laws
of nature are the same for all observers in inertial
(non-accelerated) reference frames, and none of them is preferred;
(2)\ the value of the speed of light in vacuum is a constant $c$ in
all inertial reference systems. Hence, two observers in different
inertial reference frames will be constrained to measure time
intervals and lengths in such a way that $c$ is the same constant
for both of them.

A proper time interval $\Delta t_{0}$\ is the time lapse between two
events at the \emph{same location}, as measured by a stationary
clock at that location \cite{Halliday}. The reference frame in which
one measures the proper time may be in relative motion with respect
to another inertial reference system (the reciprocal velocity being
$v$, say). In this second reference frame, the events will occur in
\emph{different places}, and the time interval $\Delta t$\ will be
\begin{equation}
\Delta t=\frac{\Delta t_{0}}{\sqrt{1-\left(  v/c\right)  ^{2}}}%
~.\label{proper time}%
\end{equation}
The Lorentz factor
\begin{equation}
\gamma=\frac{1}{\sqrt{1-\beta^{2}}}\label{Lorentz factor}%
\end{equation}
is always\ greater than $1$, once the speed parameter%
\begin{equation}
\beta=\frac{v}{c}\label{speed parameter}%
\end{equation}
is less than $1$ for any nonzero relative velocity. Therefore,
$\Delta
t>\Delta t_{0}$\ and we get the time dilation relation%
\begin{equation}
\Delta t=\gamma\Delta t_{0}~.\label{time dilation}%
\end{equation}

The length of an object measured in an inertial reference frame in
which this object is at rest is the proper length $l_{0}$. In any
other reference system in relative motion with the previous one, an
observer will measure a contracted length $l$\
\begin{equation}
l=\frac{l_{0}}{\gamma}~.\label{length contraction}%
\end{equation}
Eq. (\ref{length contraction}) is a direct consequence of time
dilation \cite{Halliday}.

With these concepts fresh in our minds, let us go back to Einstein's
thought-experiment with the elevator: and so begins the
Lenz-Sommerfeld's argument. We will be concerned with the elevator's
free fall in a gravitational field (rather than with the part of the
\textit{gedanken} experiment when we study the elevator accelerating
upwards in the absence of gravitational influence).

Consider a reference frame $K_{\infty}$\ attached to the elevator.
The other reference frame, called $K$, will be fixed at the centre
of a spherically symmetric source of the gravitational field. This
source can be taken as the Sun, of mass $M$. The elevator
$K_{\infty}$\ will be falling radially toward $K$, which may be
regarded as at rest. As we discussed in Section \ref{Sec-Intro},
such a freely falling system perceives a world free from gravitation
and, hence, from curvature. This means that an observer in
$K_{\infty}$ will measure distances according to the flat Minkowski
line
element of special relativity%
\begin{equation}
ds^{2}=dx_{\infty}^{2}+dy_{\infty}^{2}+dz_{\infty}^{2}-c^{2}dt_{\infty}%
^{2}~,\label{Minkowski}%
\end{equation}
where $\left(  x_{\infty},y_{\infty},z_{\infty},t_{\infty}\right)  $
are the coordinates measured in $K_{\infty}$.

Let $\left(  r,\theta,\phi,t\right)  $ be the coordinates measured
in system $K$ of the Sun, which is subjected to gravitation. Suppose
that the \textquotedblleft moving\textquotedblright\ system
$K_{\infty}$\ arrives at a
distance $r$ from the system $K$ at \textquotedblleft rest\textquotedblright%
\ with velocity $v=\beta c$, cf. Eq. (\ref{speed parameter}). In
addition, the $x_{\infty}$-axis will be taken as the direction of
motion: longitudinal, the same as $r$ direction. This way,
$y_{\infty}$ and $z_{\infty}$\ are transversal directions. There
will be length contraction\ along the direction of motion, meaning
that the intervals $dr$ and $dx_{\infty}$\ will be related by Eq.
(\ref{length contraction}). Two events -- such as to turn on a light
tube and then turn it off\ -- are measured at the\emph{\ same place}
in the moving reference frame $K_{\infty}$ only. Therefore, the
proper length is measured by the observer inside the elevator:
$dx_{\infty}$\ is the proper
length while $dr$\ is the contracted interval,%
\begin{equation}
dr=\frac{dx_{\infty}}{\gamma}~.\label{Lorentz contraction}%
\end{equation}
There is no contraction along the directions $y_{\infty}$ and $z_{\infty}%
$\ orthogonal to the radial motion. For this reason, these $K_{\infty}%
$-coordinates will relate to the spatial $K$-coordinates through the
simple coordinate transformations relating Cartesian coordinates and
spherical
coordinates:%
\begin{equation}
dy_{\infty}=rd\theta\label{dy}%
\end{equation}
and%
\begin{equation}
dz_{\infty}=r\sin\theta d\phi\label{dz}%
\end{equation}
Moreover, as the proper time interval separates two events happening
at the same location, it is measured in the system $K_{\infty}$.
Consequently, the dilated time interval will be the one realized in
reference frame $K$. From
Eq. (\ref{time dilation}), it results:%
\begin{equation}
dt=\gamma dt_{\infty}~.\label{Einstein dilation}%
\end{equation}

By substituting Eqs. (\ref{Lorentz contraction}) to (\ref{Einstein
dilation})
into Minkowski line element (\ref{Minkowski}), one obtains:%
\begin{equation}
ds^{2}=\frac{dr^{2}}{\left(  1-\beta^{2}\right)  }+r^{2}\left(
d\theta ^{2}+\sin^{2}\theta d\phi^{2}\right)  -c^{2}dt^{2}\left(
1-\beta^{2}\right)
~,\label{Lenz}%
\end{equation}
where we have used Eq. (\ref{Lorentz factor}). The coordinates in
line element (\ref{Lenz}) are those measured in the system of
reference $K$ attached to the Sun. Notwithstanding, the factor
$\beta$\ appearing in (\ref{Lenz}) is meaningful in connection with
the frame fixed in our elevator: only then $\beta$ can be
interpreted as the speed parameter $\beta=v/c$, because it is the
box $K_{\infty}$ that carries continuously with itself the
pseudo-Euclidean metric of special relativity. The meaning of
$\beta$\ in the reference frame $K$ is determined using
\emph{conservation of energy}.

In the approach by Lenz and Sommerfeld, the interpretation of
gravity as a long range force is combined with the one in which a
gravitational field means curvature of space and time. The metric
structure in Eq. (\ref{Minkowski}) and (\ref{Lenz}) is consistent
with this last point of view. In considering the energy of the
elevator, we will use the Newtonian ideas for gravity when assuming
that the elevator bears a \emph{potential energy} $U\left(  r\right)
$. We claim that this can be regarded as an adequate first
approximation. This potential energy is calculated in the standard
way \cite{Halliday} as
\begin{equation}
F\left(  r\right)  =-\frac{dU}{dr}\label{conservative force}%
\end{equation}
or%
\begin{equation}
U\left(  r\right)
=-\int_{r_{\infty}}^{r}dr^{\prime}~\hat{r}^{\prime
}.\mathbf{F}\left(  r^{\prime}\right)  ~,\label{integral U}%
\end{equation}
We use $r^{\prime}$\ within the integration sign to avoid confusion
with the limit of integration. (Nevertheless the meaning is clear:
either $r$ or $r^{\prime}$ denote the radial direction.) Following
Lenz, Sommerfeld substitutes the force $\mathbf{F}\left(  r\right)
$ that enters Eq.
(\ref{integral U}) by the Newton's law of gravitation, Eq. (\ref{Newton force}%
). Integrating, it results:%
\begin{equation}
U\left(  r\right)  =-\frac{GmM}{r}-U\left(  r_{\infty}\right)
\label{potential energy}%
\end{equation}
where%
\begin{equation}
U\left(  r_{\infty}\right)  =-\frac{GmM}{r_{\infty}}~.\label{U(r infinity)}%
\end{equation}
As the potential is always defined up to a constant, one may set its
zero
level at $r=r_{\infty}$:%
\begin{equation}
U\left(  r_{\infty}\right)  =0~.\label{U zero level}%
\end{equation}
This is easily justified: by taking $r_{\infty}\rightarrow\infty$,
it follows $U_{N}\left(  r_{\infty}\right)  =0$, so that
\begin{equation}
U_{N}\left(  r\right)  =-\frac{GmM}{r}~.\label{Newton potential}%
\end{equation}
(The index $N$ of $U(r)$ stands for \textit{Newtonian} potential.)

The expression for the relativistic \emph{kinetic energy} of the elevator is:%
\begin{equation}
T=m_{0}c^{2}\left(  \gamma-1\right)  =\left(  m-m_{0}\right)  c^{2}%
\label{kinetic energy}%
\end{equation}
where its mass $m$,%
\begin{equation}
m=\frac{m_{0}}{\sqrt{1-\beta^{2}}}\label{m(beta)}%
\end{equation}
is given in terms of the rest mass $m_{0}$ \cite{Feynman}. Notice
that when the elevator is at rest $v=0$, $m=m_{0}$ and consequently
$T=0$. This situation occurs as initial condition since it is
assumed that the elevator
falls from rest at $r=r_{\infty}$:%
\begin{equation}
T\left(  r_{\infty}\right)  =0~.\label{T(r infinity)}%
\end{equation}

The conservation of the \emph{total energy}
\begin{equation}
E=T+U\label{E}%
\end{equation}
gives:
\begin{equation}
E\left(  r\right)  =E\left(  r_{\infty}\right) \label{conservation of E}%
\end{equation}
However, in face of Eqs. (\ref{U zero level}) and (\ref{T(r
infinity)}), $E\left(  r_{\infty}\right)  =0$, and we are left with
\begin{equation}
\left(  m-m_{0}\right)  c^{2}-\frac{GmM}{r}=0~,\label{E = 0}%
\end{equation}
where we have used Eqs. (\ref{potential energy}) and (\ref{kinetic
energy}) for the potential and kinetic energy of the elevator at a
distance $r$ from the Sun as measured by an observer on $K$.

Dividing Eq. (\ref{E = 0}) by the mass $m$ of the test particle,%
\begin{equation}
\left(  1-\frac{m_{0}}{m}\right)  -\frac{GM}{c^{2}}\frac{1}{r}%
=0~,\label{E/m = 0}%
\end{equation}
and substituting the ratio of the masses according to Eq. (\ref{m(beta)}):%
\begin{equation}
\left(  1-\sqrt{1-\beta^{2}}\right)  -\frac{\alpha}{r}=0~.\label{beta Eq.}%
\end{equation}
$\alpha$ is the definition of the geometrical mass -- Eq.
(\ref{geometrical mass}) -- appearing in the Schwarzschild line
element. For
our Sun, $\alpha\simeq1.5$ km \cite{NASA}. From Eq. (\ref{beta Eq.}):%
\[
\sqrt{1-\beta^{2}}=1-\frac{\alpha}{r}~,
\]
i.e.,%
\begin{equation}
\left(  1-\beta^{2}\right)  =\left(  1-\frac{\alpha}{r}\right)
^{2}=\left[ 1-\frac{2\alpha}{r}+\left(  \frac{\alpha}{r}\right)
^{2}\right]
~.\label{(1-beta2)}%
\end{equation}

It is come the time for an \emph{approximation}. We will consider
that the test particle occupies positions at great distances from
the source, so that
\begin{equation}
\frac{\alpha}{r}<<1~.\label{alpha approx}%
\end{equation}
Under the approximation (\ref{alpha approx}), the term $\left(
\frac{\alpha }{r}\right)  ^{2}$ of Eq. (\ref{(1-beta2)}) is a second
order one in
$\frac{\alpha}{r}$. Hence it is negligible, leading to:%
\begin{equation}
\left(  1-\beta^{2}\right)  \simeq\left(  1-\frac{2\alpha}{r}\right)
~.\label{beta}%
\end{equation}
This determines the meaning of $\beta$ in the system $K$.

Substituting Eq. (\ref{beta}) into our expression for the line
element, Eq.
(\ref{Lenz}), we find:%
\begin{equation}
ds^{2}=\frac{dr}{\left(  1-\frac{2\alpha}{r}\right)  }+r^{2}\left(
d\theta^{2}+\sin^{2}\theta d\phi^{2}\right)  -c^{2}dt^{2}\left(
1-\frac{2\alpha}{r}\right)  ~.\label{Lenz-Schw}%
\end{equation}
This is precisely the Schwarzschild solution (\ref{Schwarzschild})
of Einstein's equation. The power of Lenz-Sommerfeld's argument is
thus unveiled: the derivation presented here could only guarantee an
approximated result; however it coincides with the exact solution of
the ten non-linear coupled partial differential equations
(\ref{Einstein Eq.}) in the presence of a massive source.

\section{The de Sitter solution via Lenz's argument \label{Sec-dS}}

Our additional step in this paper is to add to $\mathbf{F}\left(
r\right) $\ the contribution coming from the de Sitter's force, Eq.
(\ref{dS force}). Therefore, we admit that the test particle of mass
$m$ (elevator) is subjected to the linear repulsive force
$\mathbf{F}_{\Lambda}\left(  r\right)  $\ coming as an effect of a
non-null cosmological constant $\Lambda$, and which leads to a
potential
\begin{equation}
U\left(  r\right)  =-m\frac{\Lambda c^{2}}{6}r^{2}-U\left(
r_{\infty}\right)
\label{potential energy dS}%
\end{equation}
where%
\begin{equation}
U\left(  r_{\infty}\right)  =-m\frac{\Lambda c^{2}}{6}r_{\infty}%
^{2}~.\label{U(r infinity) dS}%
\end{equation}
after integration of (\ref{integral U}). Again, the additive
constant in the expression for the potential energy can be set to
zero, but the explanation for taking
\begin{equation}
U\left(  r_{\infty}\right)  =0\label{U zero level dS}%
\end{equation}
in this case associated with $\Lambda$ is different from the
Schwarzschild case. The pathology here comes from the fact that
$r_{\infty}=\infty$ would imply $U_{\Lambda}\left(
r_{\infty}\right)  \rightarrow\infty$. This apparent problem is
solved by taking $r_{\infty}=0$\ which automatically renders the
potential null $U_{\Lambda}\left(  0\right)  =0$ by Eq. (\ref{U(r
infinity) dS}). Thus, what we are really doing is assuming the level
zero for the potential at the \textquotedblleft
origin\textquotedblright\ of the de Sitter solution rather than at
its \textquotedblleft bondary\textquotedblright. Instead of
calculating the potential as the energy necessary to bring a test
particle from the infinity to the position $r$, we define
$U_{\Lambda}\left(  r\right)  $ by considering the transport of the
test particle from $r=0$ to an arbitrary position $r$.\ Therefore,%
\[
U_{\Lambda}\left(  r\right)  =-m\frac{\Lambda c^{2}}{6}r^{2}%
\]

In the de Sitter case, the equation for the relativistic
\emph{kinetic energy} of the elevator is still Eq. (\ref{kinetic
energy}). We also take the elevator at rest at the de Sitter radius
$r_{\infty}=L$, leading once more to the constraint $T\left(
r_{\infty}\right)  =0$. This initial condition is in agreement with
the predictions of general relativity: on Tolman's book
\cite{To87},\ one can read about the motion of a test particle in
the de Sitter universe and understand, based on the integration of
the geodesic equations, that it takes an infinity time for the
particle traveling toward the boundary to reach the horizon, where
its velocity would be zero.

Imposing conservation of the \emph{total energy} for the elevator's
free fall in de Sitter universe, gives now
\begin{equation}
\left(  m-m_{0}\right)  c^{2}-m\frac{\Lambda c^{2}}{6}r^{2}%
=0~.\label{E = 0 dS}%
\end{equation}
If we proceed as in Section \ref{Sec-Lenz}, and divide Eq. (\ref{E =
0 dS}) by
the mass $m$ of the test particle, we come to%
\[
\sqrt{1-\beta^{2}}=1-\frac{\Lambda}{6}r^{2}~,
\]
i.e.,%
\begin{equation}
\left(  1-\beta^{2}\right)  =\left(  1-\frac{\Lambda}{6}r^{2}\right)
^{2}=\left[  1-\frac{\Lambda}{3}r^{2}+\left(
\frac{\Lambda}{6}r^{2}\right)
^{2}\right]  ~.\label{(1-beta2) dS}%
\end{equation}

At this point we will perform an \emph{approximation}, assuming that
the value of the cosmological constant is small
\begin{equation}
\Lambda r^{2}<<1~,\label{Lambda approx}%
\end{equation}
which is actually true in view of the constraints imposed by the
dynamics of our solar system and the standard cosmological model. In
fact, one can easily
estimates $\Lambda\simeq1.38\times10^{-52}~%
\operatorname{m}%
^{-2}$\ using recent astrophysical data available at the Particle
Data Group website \cite{PDG}. The effects of the cosmological
constant are only important in a cosmological perspective, when the
repulsive force (\ref{dS force}) becomes significantly effective
(because $r$ is very large). Under the approximation (\ref{Lambda
approx}), the term $\left(  \frac {\Lambda}{6}r^{2}\right)  ^{2}$ of
Eq. (\ref{(1-beta2) dS}) is of second order
in $\Lambda r^{2}$:%
\begin{equation}
\left(  1-\beta^{2}\right)  \simeq\left(
1-\frac{\Lambda}{3}r^{2}\right)
~.\label{beta dS}%
\end{equation}
Inserting Eq. (\ref{beta dS}) into Eq. (\ref{Lenz}), renders the line element%
\begin{equation}
ds^{2}=\frac{dr}{\left(  1-\frac{\Lambda}{3}r^{2}\right)
}+r^{2}\left( d\theta^{2}+\sin^{2}\theta d\phi^{2}\right)
-c^{2}dt^{2}\left(
1-\frac{\Lambda}{3}r^{2}\right)  ~,\label{Lenz-dS}%
\end{equation}
the de Sitter solution in its static form (\ref{deSitter}). We just
got another standard exact solution of Einstein's equation from
Lenz-Sommerfeld's approximative procedure.

\section{Final comments \label{Sec-Conclusion}}

In this paper we reviewed the argument by Lenz and Sommerfeld
leading to the Schwarzschild solution of general relativity using
only concepts from the special theory of relativity and the
Newtonian theory of gravity.

We showed that it is also possible to employ Lenz-Sommerfeld's
technique to obtain a solution that includes the cosmological
constant. So we built the de Sitter solution. The later is of great
importance for modern cosmology once it allows for an accelerated
expansion of the universe, something that is favorable to the recent
data. The derivation of de Sitter solution from Lenz's argument was
ultimately possible due to the universal character of the
cosmological constant: it responds to gravitation in the same way as
all masses do.

The spirit of the paper is to point out a non-standard procedure of
deriving the classical solutions of general relativity. A natural
following step would be to apply Lenz-Sommerfeld's reasoning to
calculate the metric of a slowly rotating massive source, known as
Lenz-Thirring line element. This is an investigation under
development by the authors.

\begin{acknowledgments}
BMP thanks CNPq-Brazil for partial financial support.
\end{acknowledgments}

\end{document}